\documentclass{webofc}
\usepackage[varg]{txfonts}  
\usepackage{hyperref}
\usepackage{graphicx}
\usepackage{xcolor}

\begin{document}
\title{Bottomonium observables in an open quantum system using the quantum trajectories method}
\author{\firstname{Peter} \lastname{Vander Griend}\inst{1}\fnsep\thanks{\email{vandergriend@tum.de}}}
\institute{Physik-Department, Technische Universit\"{a}t M\"{u}nchen, James-Franck-Str. 1, 85748 Garching, Germany}

\abstract{
  We solve the Lindblad equation describing the Brownian motion of a Coulombic heavy quark-antiquark pair in a strongly coupled quark gluon plasma using the Monte Carlo wave function method. 
  The Lindblad equation has been derived in the framework of pNRQCD and fully accounts for the quantum and non-Abelian nature of the system. 
  The hydrodynamics of the plasma is realistically implemented through a 3+1D dissipative hydrodynamics code. 
  We compute the bottomonium nuclear modification factor and elliptic flow and compare with the most recent LHC data. 
  The computation does not rely on any free parameter, as it depends on two transport coefficients that have been evaluated independently in lattice QCD. Our final results, which include late-time feed down of excited states, agree well with the available data from LHC 5.02 TeV PbPb collisions.
}

\maketitle

\section{Introduction}\label{intro}
A striking feature of the strong nuclear force is its confining nature.
At low energies and macroscopic scales, the quarks and gluons charged under the strong interaction only exist in neutral configurations.
In the first few instants after the Big Bang, the energy density of the Universe was much higher than today, and the quarks and gluons, now confined in nucleons, may have propagated freely.
A recent triumph of experimental nuclear physics, the heavy ion collision experiments at the Large Hadron Collider (LHC) at CERN and the relativistic heavy ion collider (RHIC) at Brookhaven National Laboratory may have achieved sufficiently high energy densities to create a new state of matter, the quark gluon plasma (QGP), in which quarks and gluons are deconfined.
This has spurred theorists to investigate observables signaling the formation of a QGP.

One such observable is the nuclear modification factor $R_{AA}$ defined as the ratio of the yield of a heavy quarkonium state in an $AA$ collision to the yield in a $pp$ collision (adjusted for the number of binary collisions).
$R_{AA}<1$ signals heavy quarkonium suppression and has long been theorized to signal the formation of a QGP \cite{Matsui:1986dk,Karsch:1987pv}.
It was initially theorized that such a suppression would be caused by Debye screening of the inter-quark potential at distances larger than approximately the inverse of the Debye mass.
In the years since these initial predictions, our understanding of the mechanism of heavy quarkonium suppression shifted due to the finding of thermal corrections to the real part of the in-medium potential related to screening and a non-zero imaginary part related to in-medium dissociation due to Landau damping \cite{Laine:2006ns}.
This was subsequently extended by the inclusion of non-Abelian singlet-octet transitions using the effective field theory (EFT) potential non-relativistic QCD (pNRQCD) \cite{Brambilla:2008cx,Escobedo:2008sy,Brambilla:2010vq,Beraudo:2007ky}.

Due to the hierarchies of scale inherent to in-medium heavy quarkonium, EFT methods, and specifically pNRQCD, and the formalism of open quantum systems (OQS) are especially useful tools to describe the combined system.
In Refs.~\cite{Brambilla:2016wgg,Brambilla:2017zei}, pNRQCD and OQS methods were applied to in-medium bottomonium, and an in-medium evolution equation taking the form of a Lindblad equation was derived.
Refs.~\cite{Brambilla:2020qwo,Brambilla:2021wkt} made use of the \texttt{QTraj} code presented in Ref.~\cite{Omar:2021kra}, which implements the quantum trajectories algorithm, to solve the Lindblad equation and present $R_{AA}$ and the elliptic flow $v_{2}$ of the $\Upsilon(1S)$, $\Upsilon(2S)$, and $\Upsilon(3S)$ as functions of centrality and transverse momentum $p_{T}$.
The remainder of this proceeding is structured as follows: in Sec.~\ref{sec:theory_background}, we provide the necessary theory background to describe in-medium heavy quarkonium; in Sec.~\ref{sec:computational_methods}, we detail the quantum trajectories algorithm and the \texttt{QTraj} code used to solve the Lindblad equation describing the in-medium evolution of heavy quarkonium; in Sec.~\ref{sec:results}, we present our results; and we conclude in Sec.~\ref{sec:conclusions}.
We emphasize that the method and results presented in this work are fully quantum, non Abelian, heavy quark number conserving, and account for dissociation and recombination.

\section{Theory background}\label{sec:theory_background}
In this section, we present the theoretical tools, namely pNRQCD and OQS methods, necessary to describe in-medium heavy quarkonium realizing the hierarchy of scales
\begin{equation}\label{eq:hierarchy_of_scales}
	M \gg 1 / a_{0} \gg \pi T \sim m_{D} \gg E,
\end{equation}
where $M$ is the heavy quark mass, $a_{0}$ is the Bohr radius of the bound state, $T$ is the temperature of the medium, $m_{D}\sim gT$ is the Debye mass, and $E$ is the binding energy of the state.
We direct the reader to Refs.~\cite{Brambilla:2016wgg,Brambilla:2017zei} for a more detailed discussion.

pNRQCD is an EFT of the strong interaction obtained from full QCD via the sequential integrating out of the hard scale $M$ from full QCD to obtain the EFT non-relativistic QCD (NRQCD) \cite{Caswell:1985ui,Bodwin:1994jh}.
From NRQCD, further integrating out the soft scale $Mv$, where $v\ll 1$ is the relative quark-antiquark velocity in a heavy-heavy bound state, gives pNRQCD \cite{Pineda:1997bj,Brambilla:1999xf,Brambilla:2004jw}.
The degrees of freedom of the resulting theory are heavy-heavy bound states in color singlet and octet configurations and gluons and light quarks at the ultra-soft scale $Mv^{2}$.
At the Lagrangian level, pNRQCD implements a double expansion in the bound state radius $r$ and the inverse of the heavy quark mass $M$
\begin{equation}
	\mathcal{L}_{\text{pNRQCD}} = \text{Tr}\bigg[ S^{\dagger}(i\partial_{0} - h_{s}) S + O^{\dagger}(iD_{0} - h_{o}) O + O^{\dagger} \mathbf{r} \cdot g\, \mathbf{E} \, S 
	+ S^{\dagger} \mathbf{r} \cdot g\, \mathbf{E} \, O + \frac{1}{2} O^{\dagger} \left\{ \textbf{r} \cdot g\, \mathbf{E} \,,\, O \right\} \bigg],
\end{equation}
where $h_{s,o}$ are the singlet and octet Hamiltonians $h_{s,o}=\tfrac{\mathbf{p}^{2}}{M}+V_{s,o}$ with 
\begin{equation}
	V_{s} = -\frac{C_{f} \alpha_{s}(1/a_{0})}{r} ,\quad V_{o} = \frac{\alpha_{s}(1/a_{0})}{2N_{c}r},
\end{equation}
are the attractive singlet Coulombic and repulsive octet Coulombic potentials, respectively.
$C_{f}=(N_{c}^{2}-1)/(2N_{c})$ where $N_{c}=3$ is the number of colors, and $\alpha_{s}(1/a_{0})$ is the strong coupling evaluated at the inverse of the Bohr radius.
The potentials encode the information of the integrated-out soft scale.

The OQS formalism allows for the rigorous description of a quantum system of interest coupled to an environment.
Within the OQS formalism, three time scales characterize the combined system: the system intrinsic time scale $\tau_{S}$, the environment correlation time $\tau_{E}$, and the relaxation time $\tau_{R}$.
For in-medium heavy quarkonium, these scales are given by
\begin{equation}
	\tau_{S} \sim \frac{1}{E} ,\quad
	\tau_{E} \sim \frac{1}{\pi T} ,\quad
	\tau_{R} \sim \frac{1}{\Sigma_{s}} \sim \frac{1}{a_{0}^{2}(\pi T)^{3}},
\end{equation}
where $\Sigma_{s}$ is the thermal self energy of the quarkonium.

The hierarchy of scales of Eq.~(\ref{eq:hierarchy_of_scales}) implies that $\tau_{S},\,\tau_{R}\gg \tau_{E}$ and the system realizes quantum Brownian motion allowing for the use of two simplifying assumptions: the Born and Markov approximations.
The Born approximation amounts to a temporal course graining to time scales over which the quarkonium has no effect on the plasma, i.e., time scales over which any excitation caused by the presence of the quarkonium in the medium have relaxed back.
At the calculational level, this implies the density matrix of the combined quarkonium-plasma system may be taken to factorize, i.e., $\rho(t) \propto \rho_{S}(t) \otimes \rho_{E}$.
The Markov approximation implies that only the state of the quarkonium at time $t$ is necessary to describe the evolution at time $t$.
This eliminates a memory integral and gives evolution equations which are local in time.

Refs.~\cite{Brambilla:2016wgg,Brambilla:2017zei} use pNRQCD and the OQS formalism to derive the evolution equations of in-medium heavy quarkonium in a strongly coupled medium.
The evolution equations take the form of a master equation which in the limit $\pi T \gg E$ takes the form of a Lindblad equation \cite{Lindblad:1975ef,Gorini:1975nb}
\begin{equation}\label{eq:lindblad}
	\frac{d \rho(t)}{dt} = -i\left[ H, \rho(t) \right] + \sum_{n} \left( C_{i}^{n} \rho(t) C_{i}^{n\dagger} 
	- \frac{1}{2}\left\{ C_{i}^{n\dagger}C_{i}^{n}, \rho(t) \right\} \right),
\end{equation}
where 
\begin{equation}
	\rho(t) = \begin{pmatrix} \rho_{s}(t) & 0 \\ 0 & \rho_{o}(t) \end{pmatrix} ,\quad
	H =  \begin{pmatrix} h_{s} & 0 \\ 0 & h_{o} \end{pmatrix} + \frac{r^{2}}{2} \gamma \begin{pmatrix} 1 & 0 \\ 0 & \frac{N_{c}^{2}-2}{2(N_{c}^{2}-1)} \end{pmatrix},\\
\end{equation}
\begin{equation}
	C_{i}^{0} = \sqrt{\kappa} r^{i} \begin{pmatrix} 0 & \frac{1}{\sqrt{N_{c}^{2}-1}} \\ 1 & 0 \end{pmatrix} ,\quad
	C_{i}^{1} = \sqrt{\frac{(N_{c}^{2}-4)\kappa}{2(N_{c}^{2}-1)}} r^{i} \begin{pmatrix} 0 & 0 \\ 0 & 1 \end{pmatrix}.
\end{equation}
$\rho_{s,o}(t)$ are the singlet and octet density matrices.
$\kappa$ is the heavy quark momentum diffusion coefficient describing the in-medium diffusion of a single heavy quark \cite{Casalderrey-Solana:2006fio,Caron-Huot:2007rwy}, and $\gamma$ is its dispersive counterpart first identified in Refs.~\cite{Brambilla:2016wgg,Brambilla:2017zei}. 
They are defined as 
\begin{equation}
	\kappa = \frac{g^{2}}{6N_{c}} \int_{0}^{\infty}ds \left \langle \left\{ \tilde{E}^{a,i}(s,\mathbf{0}),\tilde{E}^{a,i}(0,\mathbf{0}) \right\} \right \rangle ,\quad
	\gamma = -i\frac{g^{2}}{6N_{c}} \int_{0}^{\infty}ds \left \langle \left[ \tilde{E}^{a,i}(s,\mathbf{0}),\tilde{E}^{a,i}(0,\mathbf{0}) \right] \right \rangle,
\end{equation}
where
\begin{equation}
	\tilde{E}^{a,i}(t,\mathbf{0}) = \Omega^{\dagger}(t) E^{a,i}(t,\mathbf{0}) \Omega(t) ,\quad
	\Omega(t) = \text{exp}\left[ -ig\int_{-\infty}^{t} dt' A_{0}(t',\mathbf{0}) \right].
\end{equation}
$\kappa$ and $\gamma$ are the only free parameters entering the Lindblad equation.

\section{Computational methods}\label{sec:computational_methods}
In this section, we introduce the \texttt{QTraj} code presented in Ref.~\cite{Omar:2021kra} and used in Refs.~\cite{Brambilla:2020qwo,Brambilla:2021wkt} to solve the Lindblad equation describing the in-medium evolution of heavy quarkonium.
\subsection{Quantum trajectories algorithm}\label{subsec:qtraj_algorithm}
We make use of the quantum trajectories algorithm to solve the Lindblad equation given in Eq.~(\ref{eq:lindblad}); for a general introduction to this method, see Ref.~\cite{Daley:2014fha}.
The central idea of the quantum trajectories algorithm is to split the quantum number conserving diagonal pieces of the evolution specified by the Lindblad equation from the off-diagonal pieces which alter the quantum numbers of the state.
The former are collected into an effective Hamiltonian
\begin{equation}
	H_{\text{eff}} = H - \frac{i}{2} \sum_{n} C^{n\dagger}_{i}C_{i}^{n}.
\end{equation}
The \texttt{QTraj} code presented in Ref.~\cite{Omar:2021kra} and used in Refs.~\cite{Brambilla:2020qwo,Brambilla:2021wkt} implements the quantum trajectories algorithm to solve the Lindblad equation derived in Refs.~\cite{Brambilla:2016wgg,Brambilla:2017zei} as follows:
\begin{enumerate}
	\item Initialize the wave function $|\psi(t_{0})\rangle$.
	\item Generate a random number $0<r_{1}<1$ and evolve with $H_\text{eff}$ until
	\begin{equation*}\label{eq:heff_evolution}
		|| \, e^{-i \int_{t_{0}}^{t}dt' H_\text{eff} (t')} | \psi(t_{0}) \rangle \, ||^{2} \le r_{1}.
	\end{equation*}
	Designate the first time step fulfilling the above inequality the jump time $t_{j}$; if $t_{j}<t_{f}$, where $t_{f}$ is the simulation end time, proceed to step~\ref{step:jump} to initiate a quantum jump; otherwise, end the simulation at time $t_{f}$.
	\label{step:evolution}	
	\item \label{step:jump}Perform a quantum jump:
	\begin{enumerate}
		\item If the state is in a singlet configuration, jump to the octet configuration; if the state is in an octet configuration, generate a random number $0 < r_{2} < 1$ and jump to the singlet configuration if $r_{2}<2/7$; otherwise, remain in the octet configuration.
		\item Generate a random number $0<r_{3}<1$; if $r_{3}<l/(2l+1)$, take $l\to l-1$; otherwise, take $l \to l+1$.
		\item Multiply the wave function by $r$ and normalize.
	\end{enumerate}
	\item Return to step~\ref{step:evolution}.
\end{enumerate}
Each realization of the above algorithm is a \textit{quantum trajectory}.
The average of $N$ quantum trajectories converges to the solution of the Lindblad equation as $N\to\infty$.

\subsection{Simulation details}
\subsubsection{Input parameters}
We estimate our systematic uncertainties by varying $\kappa$ and $\gamma$.
We utilize three different parametrizations of $\hat{\kappa}(T) = \kappa(T)/T^{3}$ which we denote $\hat{\kappa}_{U}(T)$, $\hat{\kappa}_{C}(T)$, and $\hat{\kappa}_{L}(T)$ corresponding to the upper, central, and lower ``fit'' curves of Fig.~13 of Ref.~\cite{Brambilla:2020siz}.
In contrast to the direct extraction of $\kappa$, we utilize three values of $\hat{\gamma} = \gamma/T^{3} = \{-3.5,-1.75,0\}$ extracted indirectly from unquenched lattice measurements of $\delta M(1S)$; this procedure is outlined in Ref.~\cite{Brambilla:2019tpt}.
Recent lattice studies \cite{Larsen:2019bwy,Shi:2021qri} favor $\delta M(1S) \simeq 0$ (and hence $\gamma = 0$) while the studies \cite{Kim:2018yhk,Aarts:2011sm} used in Ref.~\cite{Brambilla:2019tpt} favor more negative values of $\delta M(1S)$ (and hence of $\gamma$).

We take $M$ as the mass of the bottom quark from $M=m_{b}=m_{\Upsilon(1S)}/2=4.73$ GeV with $m_{\Upsilon(1S)}$ from Ref.~\cite{ParticleDataGroup:2020ssz}.
We set the value of the strong coupling $\alpha_{s}$ by solving 
\begin{equation}
	a_{0} = \frac{2}{C_{f}\alpha_{s}(1/a_{0})m_{b}},
\end{equation}
using the 1-loop, 3-flavor running of $\alpha_{s}$ with $\Lambda_{\overline{MS}}^{N_f=3}=332$ MeV \cite{Petreczky:2020tky} giving $\alpha_{s}=0.468$.

\subsubsection{Lattice parameters}
As a numerically tractable discretization of a Dirac delta function, we initialize each quantum trajectory as a Gaussian
\begin{equation}
	\psi_{\ell}(t_{0}) \propto r^{\ell} e^{-r^{2}/(ca_{0}^{2})},
\end{equation}
with $r \psi(r)$ normalized to $1$ over the 1-dimensional lattice with $c=0.2$.
We utilize a lattice of radial volume \texttt{L}$=80\text{ GeV}^{-1}$ with \texttt{NUM}$=4096$ lattice sites corresponding to a lattice spacing $a=0.0195\text{ GeV}^{-1}$; we utilize a time step of $\texttt{dt}=0.001\text{ GeV}^{-1}$.
For an analysis of the systematic error associated with the temporal and spatial discretization and finite size effects, we direct the reader to Sec.~7 of Ref.~\cite{Omar:2021kra}.

\subsubsection{Medium interaction}
Interaction with the medium is implemented by coupling to a $3+1$D dissipative relativistic hydrodynamics code using a realistic equation of state fit to lattice QCD measurements \cite{Bazavov:2013txa}. 
The hydrodynamics code uses the quasiparticle anisotropic hydrodynamics (aHydroQP) framework \cite{Alqahtani:2015qja,Alqahtani:2016rth,Alqahtani:2017mhy} and in Ref.~\cite{Alqahtani:2020paa} was tuned to soft hadronic data from 5.02 TeV collisions using smooth optical Glauber initial conditions.
In Ref.~\cite{Brambilla:2020qwo}, per centrality bin, approximately 132000 trajectories through the plasma were Monte Carlo sampled and the temperature evolution averaged in each bin.
The Lindblad equation was solved using the single average temperature evolution in each centrality bin.
In Ref.~\cite{Brambilla:2021wkt}, approximately $7-9\times 10^{5}$ physical trajectories through the plasma were generated by Monte Carlo sampling.
For each physical trajectory, the production point was sampled in the transverse plane using the nuclear binary collision overlap profile $N_{AA}^{bin}(x,y,b)$, the initial $p_{T}$ from an $E_{T}^{-4}$ spectrum, and the azimuthal angle $\phi$ uniformly in $[0,2\pi]$.
Approximately $50-100$ quantum trajectories were generated per physical trajectory.
The distinct physical trajectories allow for the extraction of differential observables including $v_{2}$ and the presentation of results as a function of $p_{T}$.
The wave function is initialized at $t=0$ fm and evolved in the vacuum until $t=0.6$ fm at which time interaction with the medium is initialized; the wave function is evolved in the medium until the local temperature falls below $T_{f}=250$ MeV, from which time forward vacuum evolution is again used, to ensure that the hierarchy of scales of Eq.~(\ref{eq:hierarchy_of_scales}) is fulfilled at all times. 

\subsubsection{Feed down of excited states}
Using the quantum trajectories algorithm as implemented in the \texttt{QTraj} code \cite{Omar:2021kra}, one can calculate the survival probability of a quarkonium state, i.e., the probability a state traverses the QGP without dissociating. 
In order to compare with experiment, we must take into account the probability that an excited state decays into a lower-lying state after exiting the plasma.
The experimentally observed production cross section of a state is related to the direct production cross section by $\vec{\sigma}_{\text{exp}} = F \vec{\sigma}_{\text{direct}}$ where the vectors contain the states and $F$ is a matrix describing the feed down.
We take into account the states $\{\Upsilon(1S),\Upsilon(2S),\chi_{b0}(1P),\chi_{b1}(1P),\chi_{b2}(1P),\Upsilon(3S),\chi_{b0}(2P),\chi_{b1}(2P),\chi_{b2}(2P)\}$.
$F_{ij}$ where $i<j$ is the branching ratio of the state j to i, the diagonal of $F$ is $1$, and $F_{ij}=0$ for $i>j$.
$R_{AA}$ of state $i$ is calculated from the relation
\begin{equation}
	R^{i}_{AA}(c,p_{T},\phi) = \frac{\left(F \cdot S(c,p_{T},\phi) \cdot \vec{\sigma}_{\text{direct}}\right)^{i} }{\vec{\sigma}_{\text{exp}}},
\end{equation}
where $S(c,p_{T},\phi)$ is a diagonal matrix containing the survival probabilities and $c$ labels the centrality class.
The feed down fractions of $F$ are taken from the Particle Data Group \cite{ParticleDataGroup:2020ssz}.
Our experimental cross sections are $\vec{\sigma}_{\text{exp}} = \{57.6, 19, 3.72, 13.69, 16.1, 6.8, 3.27, 12.0, 14.15\}$ nb and are computed using data from the experimental measurements of Refs.~\cite{CMS:2018zza,LHCb:2014ngh} as explained in Sec.~6.4 of Ref.~\cite{Brambilla:2020qwo}.

\section{Results}\label{sec:results}
\begin{figure*}
	\begin{center}
		\includegraphics[width=0.43\linewidth]{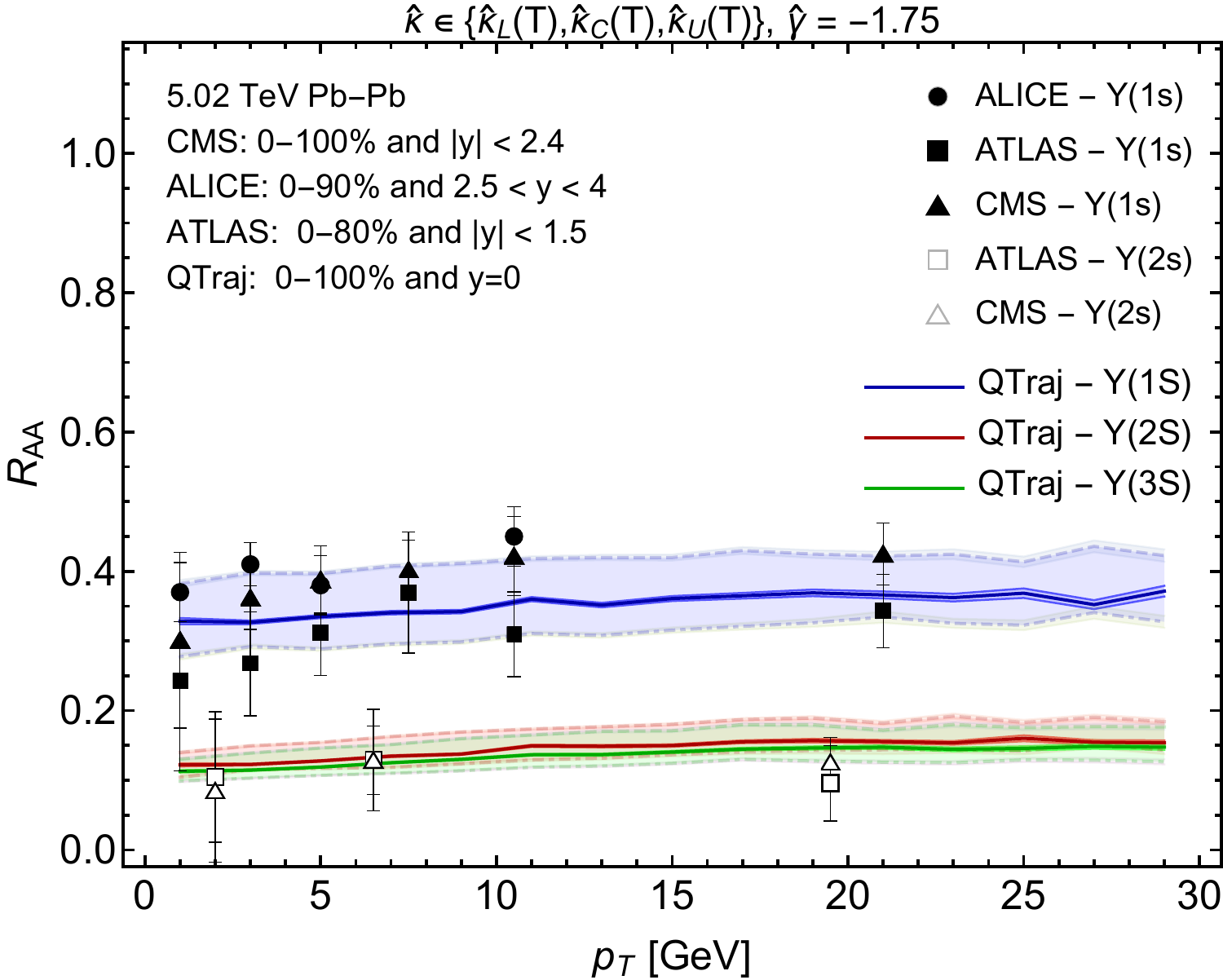}  \hspace{1cm}
		\includegraphics[width=0.43\linewidth]{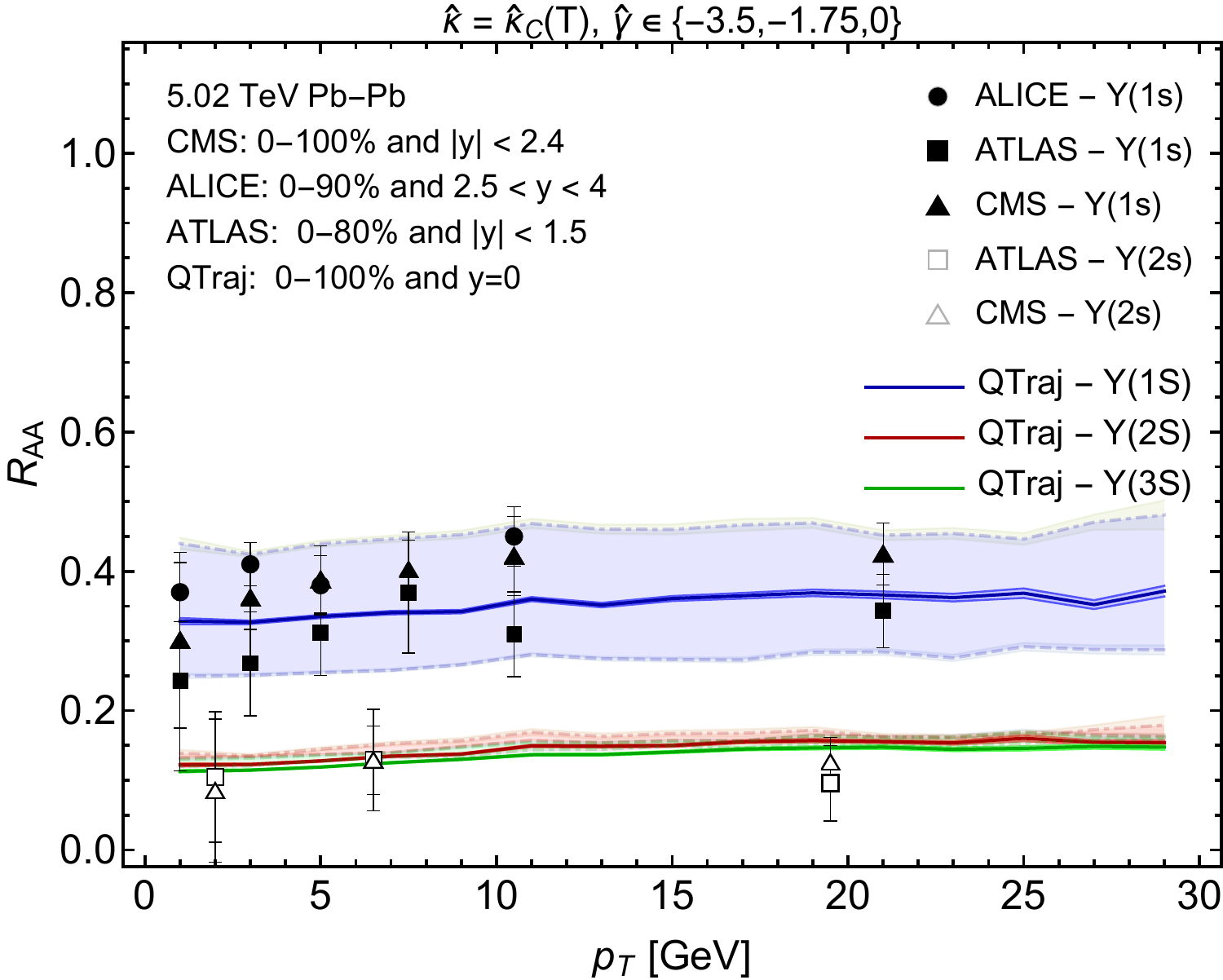}
	\end{center}
	\caption{Nuclear modification factor $R_{AA}$ of the $\Upsilon(1S)$, $\Upsilon(2S)$, and $\Upsilon(3S)$ as a function of $p_{T}$ compared to experimental measurements.
		The experimental data are taken from the ALICE~\cite{ALICE:2020wwx}, ATLAS~\cite{ATLAS5TeV}, and CMS~\cite{CMS:2018zza} collaborations.
		The bands in the theoretical curves indicate variation with respect to $\hat{\kappa}(T)$ (left) and $\hat{\gamma}$ (right).
		The central curves represent the central values of $\hat{\kappa}(T)$ and $\hat{\gamma}$, and the dashed and dot-dashed lines represent the lower and upper values, respectively, of $\hat{\kappa}(T)$ and $\hat{\gamma}$.
		Taken from Ref.~\cite{Brambilla:2021wkt}.
	}
	\label{fig:raa_vs_pt}
\end{figure*}

\begin{figure*}
	\begin{center}
		\includegraphics[width=0.44\linewidth]{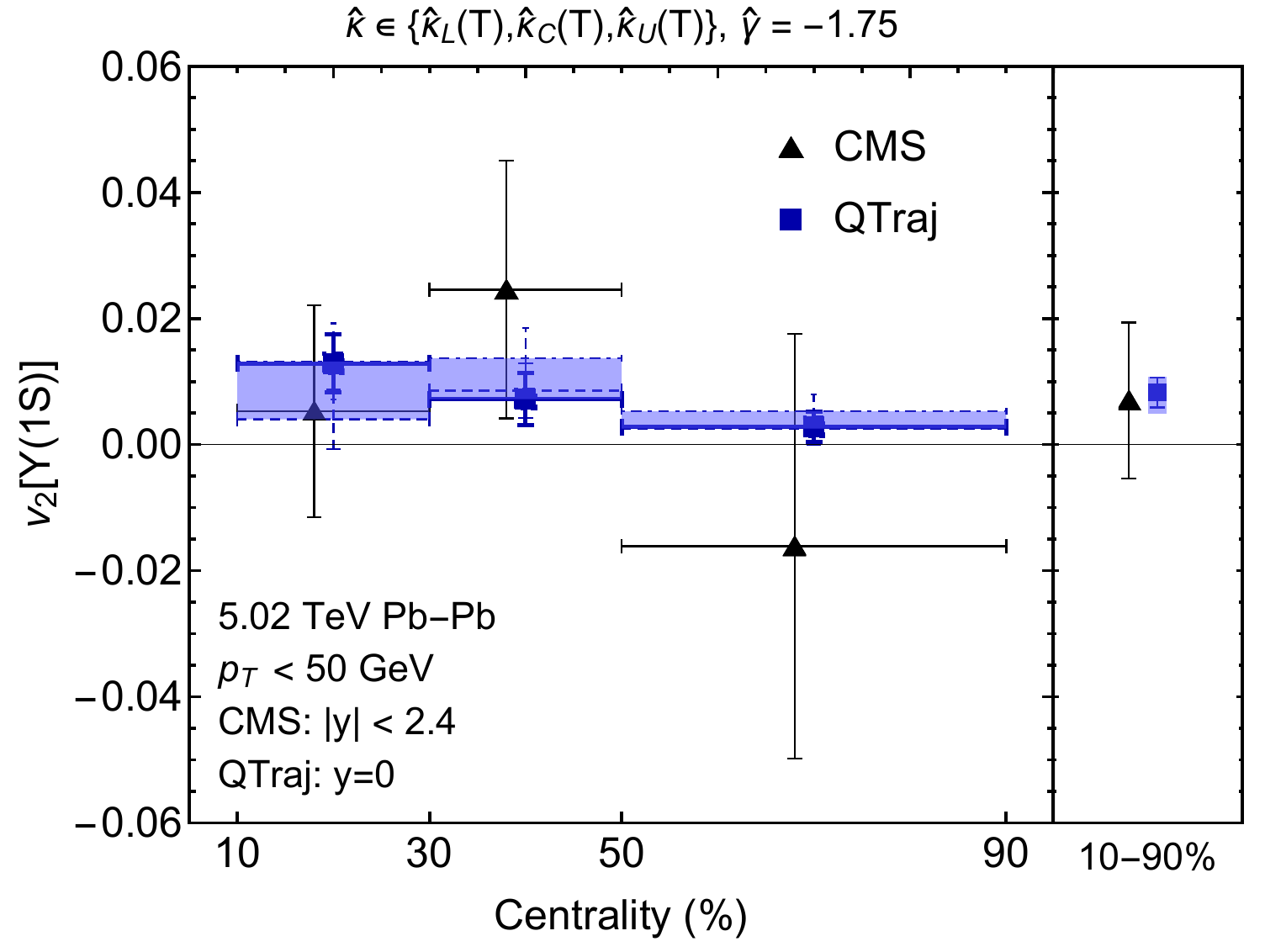} \hspace{5mm}
		\includegraphics[width=0.44\linewidth]{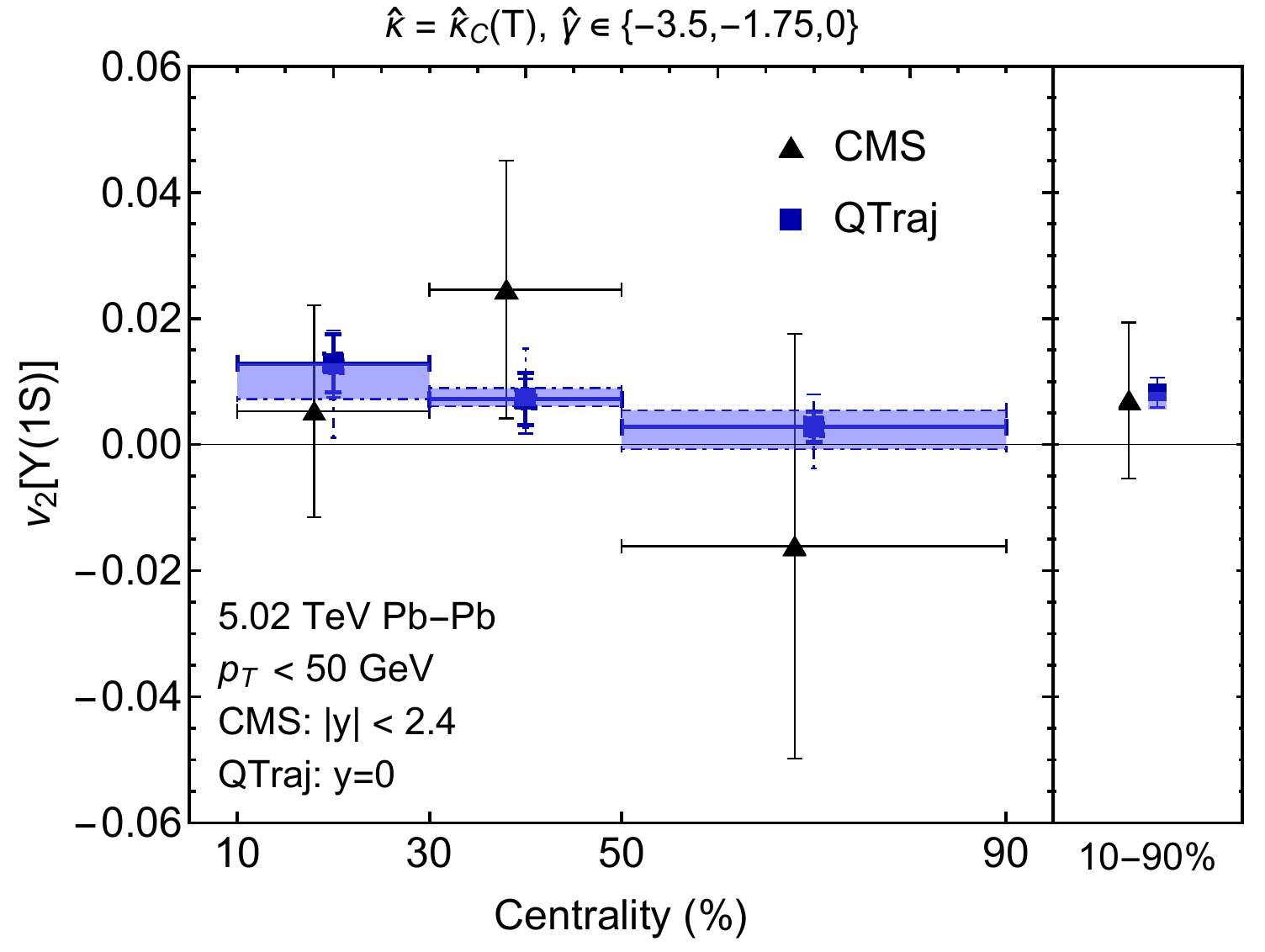}\;\;\;\;
	\end{center}
	\caption{Elliptic flow $v_{2}$ of the $\Upsilon(1S)$ as a function of centrality compared to experimental measurements of the CMS~\cite{CMS:2020efs} collaboration.
	The bands represent uncertainties as in Fig.~\ref{fig:raa_vs_pt}.
	Taken from Ref.~\cite{Brambilla:2021wkt}.
	}
	\label{fig:v2_1S_vs_centrality}
\end{figure*}

\begin{figure*}
	\begin{center}
		\includegraphics[width=0.42\linewidth]{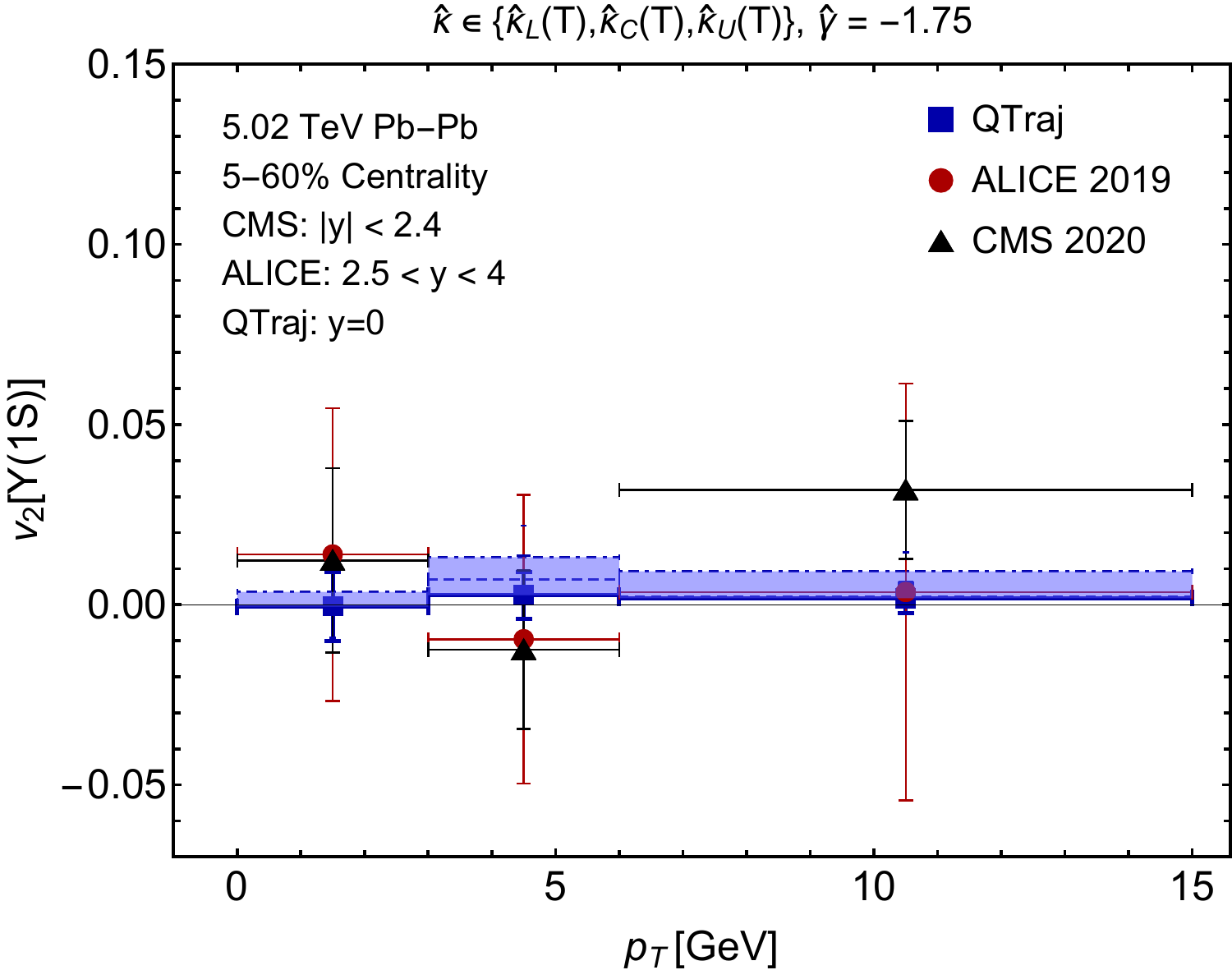} \hspace{1cm}
		\includegraphics[width=0.42\linewidth]{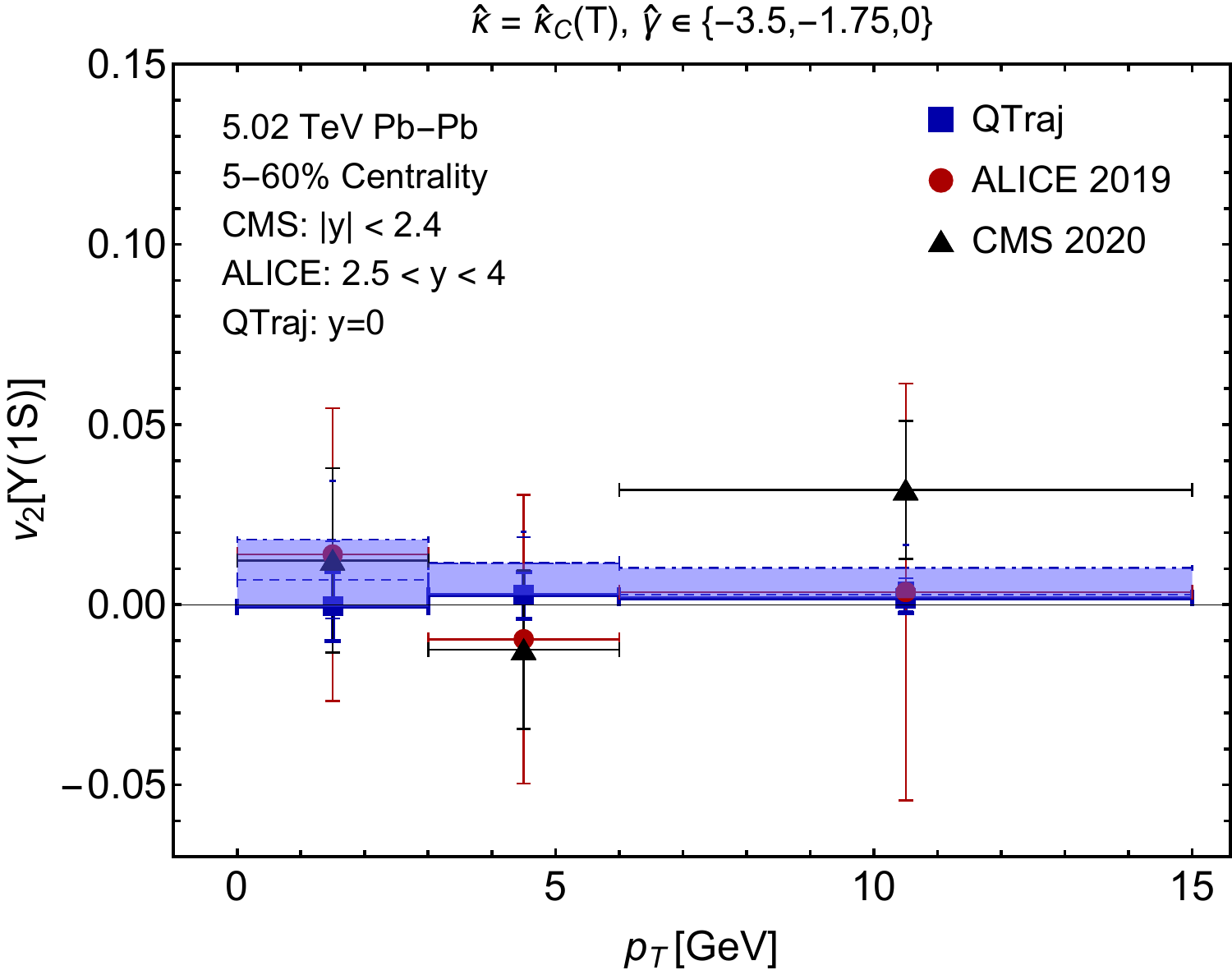}
	\end{center}
	\caption{Elliptic flow $v_{2}$ of the $\Upsilon(1S)$ as a function of $p_{T}$ compared to experimental measurements of the ALICE~\cite{ALICE:2019pox} and CMS~\cite{CMS:2020efs} collaborations.
		The bands represent uncertainties as in Fig.~\ref{fig:raa_vs_pt}.
		Taken from Ref.~\cite{Brambilla:2021wkt}.
	}
	\label{fig:v2_1S_vs_pt}
\end{figure*}

In Fig.~\ref{fig:raa_vs_pt}, we present our \texttt{QTraj} results for $R_{AA}$ of the $\Upsilon(1S)$, $\Upsilon(2S)$, and $\Upsilon(3S)$ as functions of $p_{T}$ against experimental measurements of the ALICE~\cite{ALICE:2020wwx}, ATLAS~\cite{ATLAS5TeV}, and CMS~\cite{CMS:2018zza} collaborations.
We observe good agreement with the experimental data and note the generally flat dependence of $R_{AA}$ on $p_{T}$ of both our theoretical and the experimental results.
In Figs.~\ref{fig:v2_1S_vs_centrality} and \ref{fig:v2_1S_vs_pt}, we present our \texttt{QTraj} results for $v_{2}$ of the $\Upsilon(1S)$ as a function of centrality and of $p_{T}$, respectively, against measurements of the ALICE~\cite{ALICE:2019pox} and CMS~\cite{CMS:2020efs} collaborations.
We note that our results lie within the uncertainty bounds of the experimental measurements.

In Ref.~\cite{Brambilla:2021wkt}, we present results for $R_{AA}$ of the $\Upsilon(1S)$, $\Upsilon(2S)$, and $\Upsilon(3S)$ as functions of centrality.
We additionally plot the double ratios of $R_{AA}[\Upsilon(2S)]$ and $R_{AA}[\Upsilon(3S)]$ to $R_{AA}[\Upsilon(1S)]$ against centrality and $p_{T}$.
We also give the elliptic flow of the $\Upsilon(2S)$ and $\Upsilon(3S)$ against centrality.

\section{Conclusions}\label{sec:conclusions}
The hierarchy of scales inherent to in-medium bottomonium make the EFT pNRQCD and the OQS formalism natural tools to describe the system and extract observables of interest.
Carrying out this analysis, the authors of Refs.~\cite{Brambilla:2016wgg,Brambilla:2017zei} found that the in-medium evolution equations of bottomonium take the form of a master equation which in the limit $\pi T \gg E$ takes the form of a Lindblad equation.
In Refs.~\cite{Brambilla:2020qwo,Brambilla:2021wkt}, the \texttt{QTraj} code \cite{Omar:2021kra} was utilized to solve the Lindblad equation and extract the nuclear modification factor $R_{AA}$ and the elliptic flow $v_{2}$ of the $\Upsilon(1S)$, $\Upsilon(2S)$, and $\Upsilon(3S)$ as functions of centrality and transverse momentum $p_{T}$.
The method is fully quantum, non-abelian, heavy quark number conserving, and accounts for dissociation and quantum recombination; furthermore, the $R_{AA}$ and $v_{2}$ results show good agreement with experimental measurements of the ALICE, ATLAS, and CMS collaborations.

\bibliography{bibliography} 

\begin{thebibliography}{41}

\bibitem{Matsui:1986dk}
T.~Matsui, H.~Satz, Phys. Lett. B \textbf{178}, 416 (1986)

\bibitem{Karsch:1987pv}
F.~Karsch, M.T. Mehr, H.~Satz, Z. Phys. C \textbf{37}, 617 (1988)

\bibitem{Laine:2006ns}
M.~Laine, O.~Philipsen, P.~Romatschke, M.~Tassler, JHEP \textbf{03}, 054
  (2007), \texttt{hep-ph/0611300}

\bibitem{Brambilla:2008cx}
N.~Brambilla, J.~Ghiglieri, A.~Vairo, P.~Petreczky, Phys. Rev. D \textbf{78},
  014017 (2008), \texttt{0804.0993}

\bibitem{Escobedo:2008sy}
M.A. Escobedo, J.~Soto, Phys. Rev. A \textbf{78}, 032520 (2008),
  \texttt{0804.0691}

\bibitem{Brambilla:2010vq}
N.~Brambilla, M.A. Escobedo, J.~Ghiglieri, J.~Soto, A.~Vairo, JHEP \textbf{09},
  038 (2010), \texttt{1007.4156}

\bibitem{Beraudo:2007ky}
A.~Beraudo, J.P. Blaizot, C.~Ratti, Nucl. Phys. A \textbf{806}, 312 (2008),
  \texttt{0712.4394}

\bibitem{Brambilla:2016wgg}
N.~Brambilla, M.A. Escobedo, J.~Soto, A.~Vairo, Phys. Rev. D \textbf{96},
  034021 (2017), \texttt{1612.07248}

\bibitem{Brambilla:2017zei}
N.~Brambilla, M.A. Escobedo, J.~Soto, A.~Vairo, Phys. Rev. D \textbf{97},
  074009 (2018), \texttt{1711.04515}

\bibitem{Brambilla:2020qwo}
N.~Brambilla, M.A. Escobedo, M.~Strickland, A.~Vairo, P.~Vander~Griend, J.H.
  Weber, JHEP \textbf{05}, 136 (2021), \texttt{2012.01240}

\bibitem{Brambilla:2021wkt}
N.~Brambilla, M.A. Escobedo, M.~Strickland, A.~Vairo, P.~Vander~Griend, J.H.
  Weber (2021), \texttt{2107.06222}

\bibitem{Omar:2021kra}
H.B. Omar, M.A. Escobedo, A.~Islam, M.~Strickland, S.~Thapa, P.~Vander~Griend,
  J.H. Weber (2021), \texttt{2107.06147}

\bibitem{Caswell:1985ui}
W.E. Caswell, G.P. Lepage, Phys. Lett. B \textbf{167}, 437 (1986)

\bibitem{Bodwin:1994jh}
G.T. Bodwin, E.~Braaten, G.P. Lepage, Phys. Rev. D \textbf{51}, 1125 (1995),
  [Erratum: Phys.Rev.D 55, 5853 (1997)], \texttt{hep-ph/9407339}

\bibitem{Pineda:1997bj}
A.~Pineda, J.~Soto, Nucl. Phys. B Proc. Suppl. \textbf{64}, 428 (1998),
  \texttt{hep-ph/9707481}

\bibitem{Brambilla:1999xf}
N.~Brambilla, A.~Pineda, J.~Soto, A.~Vairo, Nucl. Phys. B \textbf{566}, 275
  (2000), \texttt{hep-ph/9907240}

\bibitem{Brambilla:2004jw}
N.~Brambilla, A.~Pineda, J.~Soto, A.~Vairo, Rev. Mod. Phys. \textbf{77}, 1423
  (2005), \texttt{hep-ph/0410047}

\bibitem{Lindblad:1975ef}
G.~Lindblad, Commun. Math. Phys. \textbf{48}, 119 (1976)

\bibitem{Gorini:1975nb}
V.~Gorini, A.~Kossakowski, E.C.G. Sudarshan, J. Math. Phys. \textbf{17}, 821
  (1976)

\bibitem{Casalderrey-Solana:2006fio}
J.~Casalderrey-Solana, D.~Teaney, Phys. Rev. D \textbf{74}, 085012 (2006),
  \texttt{hep-ph/0605199}

\bibitem{Caron-Huot:2007rwy}
S.~Caron-Huot, G.D. Moore, Phys. Rev. Lett. \textbf{100}, 052301 (2008),
  \texttt{0708.4232}

\bibitem{Daley:2014fha}
A.J. Daley, Adv. Phys. \textbf{63}, 77 (2014), \texttt{1405.6694}

\bibitem{Brambilla:2020siz}
N.~Brambilla, V.~Leino, P.~Petreczky, A.~Vairo, Phys. Rev. D \textbf{102},
  074503 (2020), \texttt{2007.10078}

\bibitem{Brambilla:2019tpt}
N.~Brambilla, M.A. Escobedo, A.~Vairo, P.~Vander~Griend, Phys. Rev. D
  \textbf{100}, 054025 (2019), \texttt{1903.08063}

\bibitem{Larsen:2019bwy}
R.~Larsen, S.~Meinel, S.~Mukherjee, P.~Petreczky, Phys. Rev. D \textbf{100},
  074506 (2019), \texttt{1908.08437}

\bibitem{Shi:2021qri}
S.~Shi, K.~Zhou, J.~Zhao, S.~Mukherjee, P.~Zhuang (2021), \texttt{2105.07862}

\bibitem{Kim:2018yhk}
S.~Kim, P.~Petreczky, A.~Rothkopf, JHEP \textbf{11}, 088 (2018),
  \texttt{1808.08781}

\bibitem{Aarts:2011sm}
G.~Aarts, C.~Allton, S.~Kim, M.P. Lombardo, M.B. Oktay, S.M. Ryan, D.K.
  Sinclair, J.I. Skullerud, JHEP \textbf{11}, 103 (2011), \texttt{1109.4496}

\bibitem{ParticleDataGroup:2020ssz}
P.A. Zyla et~al. (Particle Data Group), PTEP \textbf{2020}, 083C01 (2020)

\bibitem{Petreczky:2020tky}
P.~Petreczky, J.H. Weber (2020), \texttt{2012.06193}

\bibitem{Bazavov:2013txa}
A.~Bazavov, J. Phys. Conf. Ser. \textbf{446}, 012011 (2013), \texttt{1303.6294}

\bibitem{Alqahtani:2015qja}
M.~Alqahtani, M.~Nopoush, M.~Strickland, Phys. Rev. C \textbf{92}, 054910
  (2015), \texttt{1509.02913}

\bibitem{Alqahtani:2016rth}
M.~Alqahtani, M.~Nopoush, M.~Strickland, Phys. Rev. C \textbf{95}, 034906
  (2017), \texttt{1605.02101}

\bibitem{Alqahtani:2017mhy}
M.~Alqahtani, M.~Nopoush, M.~Strickland, Prog. Part. Nucl. Phys. \textbf{101},
  204 (2018), \texttt{1712.03282}

\bibitem{Alqahtani:2020paa}
M.~Alqahtani, M.~Strickland (2020), \texttt{2008.07657}

\bibitem{CMS:2018zza}
A.M. Sirunyan et~al. (CMS), Phys. Lett. B \textbf{790}, 270 (2019),
  \texttt{1805.09215}

\bibitem{LHCb:2014ngh}
R.~Aaij et~al. (LHCb), Eur. Phys. J. C \textbf{74}, 3092 (2014),
  \texttt{1407.7734}

\bibitem{ALICE:2020wwx}
S.~Acharya et~al. (ALICE), Phys. Lett. B \textbf{822}, 136579 (2021),
  \texttt{2011.05758}

\bibitem{ATLAS5TeV}
{Songkyo Lee (ATLAS Collaboration)}, \emph{{Quarkonium production in Pb+Pb
  collisions with ATLAS}}, Quark Matter 2020
  \url{https://indico.cern.ch/event/792436/contributions/3535775/} (2017)

\bibitem{ALICE:2019pox}
S.~Acharya et~al. (ALICE), Phys. Rev. Lett. \textbf{123}, 192301 (2019),
  \texttt{1907.03169}

\bibitem{CMS:2020efs}
A.M. Sirunyan et~al. (CMS), Phys. Lett. B \textbf{819}, 136385 (2021),
  \texttt{2006.07707}

\end{thebibliography}
\bibliographystyle{woc}

\end{document}